\documentclass[iop,useAMS]{emulateapj}
\usepackage[fleqn]{amsmath}

\usepackage{graphicx}
\usepackage[percent]{overpic}
\usepackage{multirow}
\usepackage{color}
\usepackage{rotating}

\def\ltsima{$\; \buildrel < \over \sim \;$}
\def\simlt{\lower.5ex\hbox{\ltsima}}
\def\gtsima{$\; \buildrel > \over \sim \;$}
\def\simgt{\lower.5ex\hbox{\gtsima}}
\def\kms{{\rm\,km\,s^{-1}}}
\def\pc{{\rm\,pc}}
\def\kpc{{\rm\,kpc}}

\newcommand{\fmmm}[1]{\mbox{$#1$}}
\newcommand{\scnd}{\mbox{\fmmm{''}\hskip-0.3em .}}


\def\deg{^\circ}
\def\degg{\hbox{$\null^\circ$\hskip-3pt .}}
\def\sec{\hbox{"\hskip-3pt .}}

\def\Gyr{{\rm\,Gyr}}

\def\ltsima{$\; \buildrel < \over \sim \;$}
\def\gtsima{$\; \buildrel > \over \sim \;$}

\def\Pal{Palomar~5}

\shorttitle{The Palomar 5 stellar stream}
\shortauthors{Ibata et al.}

\begin{document}

\title{Feeling the pull,\\ a study of natural Galactic accelerometers. I:\\
photometry of the delicate stellar stream of the Palomar 5 globular cluster\altaffilmark{1}}

\author{Rodrigo A. Ibata\altaffilmark{2}}
\author{Geraint F. Lewis\altaffilmark{3}}
\author{Nicolas F. Martin\altaffilmark{2,4}}

\altaffiltext{1}{Based on observations obtained with MegaPrime/MegaCam, a joint project of CFHT and CEA/DAPNIA, at the Canada-France-Hawaii Telescope (CFHT) which is operated by the National Research Council (NRC) of Canada, the Institute National des Sciences de l'Univers of the Centre National de la Recherche Scientifique of France, and the University of Hawaii.}

\altaffiltext{2}{Observatoire astronomique de Strasbourg, Universit\'e de Strasbourg, CNRS, UMR 7550, 11 rue de lÕUniversit\'e, F-67000 Strasbourg, France; rodrigo.ibata@astro.unistra.fr}
\altaffiltext{3}{Sydney Institute for Astronomy, School of Physics, A28, The University of Sydney, NSW, 2006, Australia}
\altaffiltext{4}{Max-Planck-Institut f\"ur Astronomie, K\"onigstuhl 17, D-69117 Heidelberg, Germany}

\begin{abstract}
We present an analysis of wide-field photometric surveys of the Palomar 5 globular cluster and its stellar stream, based on ${\rm g}$- and ${\rm r}$-band measures together with narrow-band DDO51 photometry. In this first study, we use the deep ${\rm (g,r)}$ data to measure the incidence of gaps and peaks along the stream. 
 Examining the star-counts profile of the stream plus contaminating populations, we find no evidence for significant under-densities, and find only a single significant over-density. This is at odds with earlier studies based on matched-filter maps derived from shallower SDSS data if the contaminating population possesses plausible spatial properties.
The lack of substantial sub-structure along the stream may be used in future dynamical simulations to examine the incidence of dark matter sub-halos in the Galactic halo. We also present a measurement of the relative distances along the stream which we use to create the deepest wide-field map of this system to date.
\end{abstract}


\section{Introduction}
\label{sec:Introduction}

A natural consequence of the colossal masses of giant galaxies is that they deform and disrupt their satellite companions. Elementary considerations show that tides become important when the density of the satellite drops below the mean density of the host inside the satellite's orbit. This means that many dwarf galaxies and globular clusters at the present day are significantly affected by tidal forces, given their observed extent and inferred masses. Over time, the host galaxy grows in mass, while the satellite decays in its orbit (due to dynamical friction) and simultaneously loses mass (via the normal secular evolution processes of post-formation stellar escape, violent relaxation, three-body encounters, and slow evaporation). All of these processes render the satellite less resilient to the ever-growing tidal field of the host. Eventually, many, if not most, satellites will find themselves at a location in their host galaxy where their mass is no longer sufficient to keep the outermost stars bound, and the process of tidal dissolution begins.

The idealized situation where the satellite moves in a circular orbit, and loses stars slowly has been treated in detail by \citet{1997MNRAS.284..811R} and \citet{2000MNRAS.318..753F}. It is fascinating to find that in this situation stars find it very difficult to depart from the neighborhood of the satellite even when they have energies significantly above the escape velocity: the stars tend to ``bounce around'' inside the satellite's potential well until they eventually manage to find the Lagrange points L1 or L2, through which they can escape. Stars that escape through the L1 point find themselves lower in their host's potential, gain larger orbital velocity, and hence form the leading arm of the tidal stream. In contrast those stars that leave via the L2 point attain lower velocity and form the trailing arm. While this circular orbit configuration is clearly unlikely to happen often in nature, it nevertheless captures many of the evolutionary properties seen in N-body simulations of the formation of tidal tails from satellites on more general orbits.

\begin{figure*}
\begin{center}
\includegraphics[angle=0, viewport= 25 30 760 585, clip, width=\hsize]{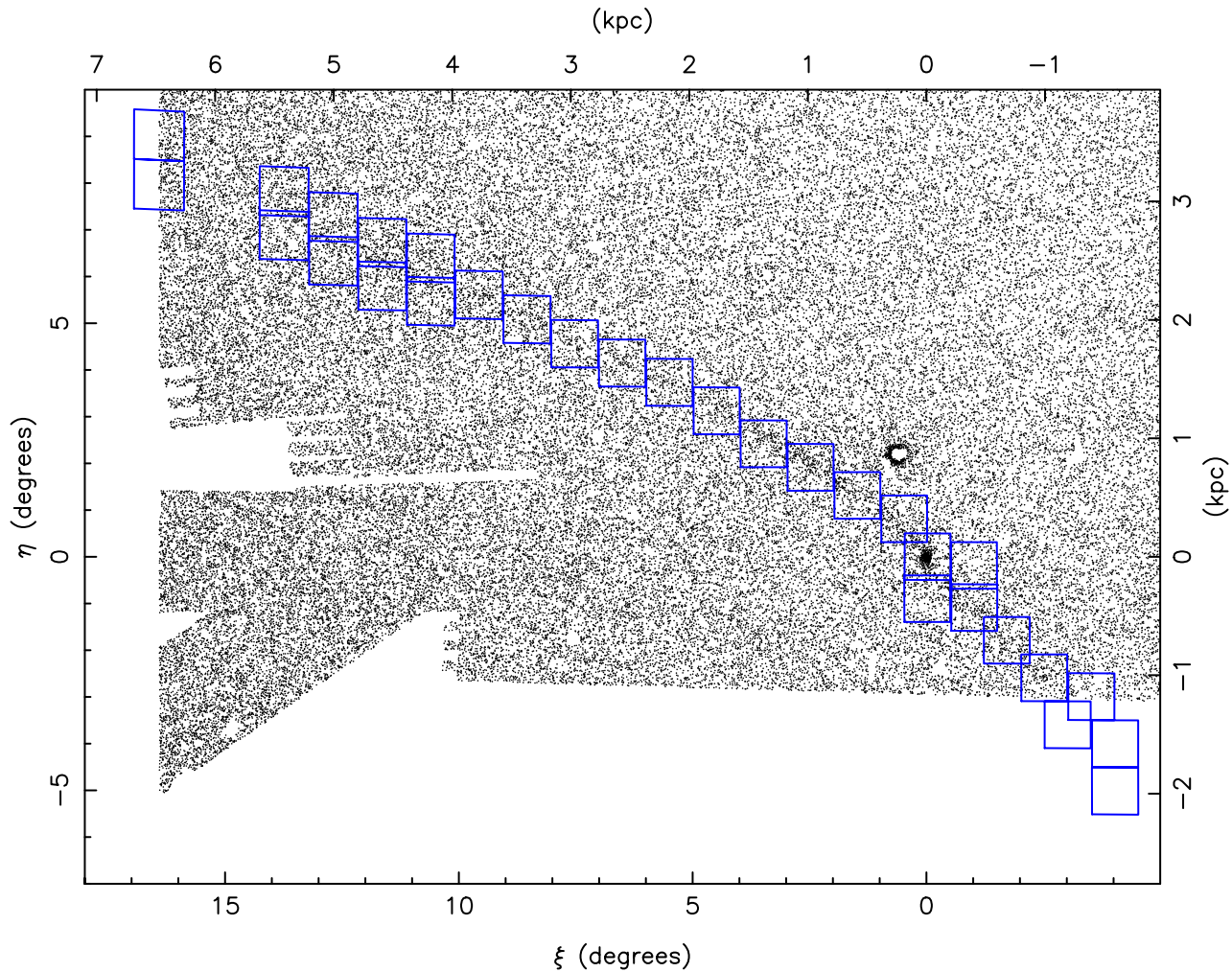}
\end{center}
\caption{Main photometric survey fields used in this study. In all the maps presented in this study, $\xi,\eta$ are standard coordinates with respect to the center of \Pal\ (i.e., $\xi,\eta$ are locally parallel to celestial coordinates at the position of \Pal, taken as the tangent point). The background black dots show the point sources in the SDSS in the vicinity of \Pal, selected from the de-reddened $(g-r)_0,g_0$ color-magnitude diagram (CMD). These stars have magnitudes $20 < g_0 < 21.5$, and were selected in a narrow CMD region around the observed turnoff of \Pal. This is close to the best map that can be obtained from the SDSS without resorting to statistical methods (that generally replace stars by weights). The \Pal\ stream lies towards the edge of the SDSS footprint, which explains the irregular coverage towards the South and East in this diagram. The 30 blue squares show the longer exposure CFHT MegaCam field pointings used in this study. For clarity we have not shown the positions of the shorter exposure fields, but these form an interlocking pattern with the longer exposure fields. The physical scale of the stream can be appreciated from the top and right-hand axes (where we have assumed a distance of $23.5\kpc$, \citealt{2011ApJ...738...74D}). The cluster remnant is clearly visible at $(\xi,\eta)=(0,0)$, while the overdensity of stars at $(\xi,\eta) \approx (+0.5,2)$ is due to the (unrelated) globular cluster M~5.}
\label{fig:photometric_fields}
\end{figure*}

It is now recognized that tidal streams and other debris hold valuable clues to the formation history of their host galaxies \citep{Johnston:2008jp}. Since mixing times are long outside of the inner $\sim 10\kpc$, stellar streams retain their structural coherence and are identifiable for many $\Gyr$ after the final dissolution of their progenitor. This therefore gives us a means to constrain the number, nature, composition and orbit of the satellites that were accreted onto their hosts. Significant effort has been devoted to the detection of such structures, especially around the giant Local Group galaxies, the Milky Way (e.g., \citealt{Ibata:2001be,2006ApJ...642L.137B,Grillmair:2006gh,2009ApJ...693.1118G,2013ApJ...762....6S, Bernard:2014ip}) and M31 \citep{2001Natur.412...49I,2009Natur.461...66M,2014ApJ...780..128I}, as well as further afield \citep{2010AJ....140..962M,Mouhcine:2010cz}. Many further discoveries are eagerly expected in the coming years thanks to new surveys in progress (PanSTARRS, LSST and particularly Gaia).

Apart from their use as fossils from the epoch of the formation of a galaxy, stellar streams also hold powerful information about the gravitational potential they inhabit. At present, the best means to determine the mass distribution of a spiral galaxy is to measure the \ion{H}{1} rotation curve, but while the resulting constraints are of excellent quality, they are spatially limited to the (highly flattened) disk, and hence probe a virtually insignificant volume of the dark halo that surrounds the galaxy. Other methods assume dynamical equilibrium in the population of satellites (e.g. via the Jeans equation, \citealt{Battaglia:2005jx}) and use the measured velocities of satellite galaxies and globular clusters, to constrain the dark halo. However, of the few satellites that are known around the giant spirals in the Local Group, many appear to be kinematically correlated \citep{2013MNRAS.435.2116P,2013Natur.493...62I}, which seriously calls into question the validity of the assumptions underlying such methods.

Stellar streams allow us to circumvent these issues. Unlike \ion{H}{1} rotation curve analyses, they are found at a range of distances, all the way into the outermost reaches of halos \citep{2014ApJ...780..128I}. Furthermore, we do not need to assume dynamical equilibrium of a tracer population because we can use the spatial and kinematic continuity along the stellar stream to probe the potential. The principle behind the method is that a low-mass stream follows closely the guiding centre orbit, so to first approximation the stars along such a stream trace out an orbit in the galaxy. Since energy
\begin{equation}
E = {{1}\over{2}} ( v_x^2 + v_y^2 +v_z^2) + \Phi
\end{equation}
is constant along an orbit, it follows that contours of $\vec{v}$ along the stream are simply contours of the potential $\Phi$. This then provides a means to measure the potential in the spatial region inhabited by the stream. While we do not generally have access to the full three-dimensional velocity vector (though note that this will change once the Gaia mission is complete), we have shown that projected velocities are often sufficient to constrain the galactic mass distribution if certain symmetry assumptions are made \citep{2011MNRAS.417..198V}. In reality streams do not form from progenitors of zero mass, and the self-gravity of the dissolving satellite alters the path of the stream stars. The algorithm (see also \citealt{2012MNRAS.420.2700K,2015MNRAS.449.1391B}) takes this self-gravity into account, correcting the guiding centre orbit to predict the stellar stream locus. The system parameters (host mass distribution, progenitor mass and orbit) are fitted using Markov Chain Monte Carlo methods. 
The particular form of the shape of the dark matter halo can also leave dynamical traces on streams; indeed it has recently been argued that the \Pal\ stellar stream excludes significantly triaxial distributions within the Galactic radial range that this stream explores \citep{2015ApJ...799...28P}.

\begin{figure}
\begin{center}
\hbox{
\includegraphics[angle=0, viewport= 45 30 775 580, clip, width=\hsize]{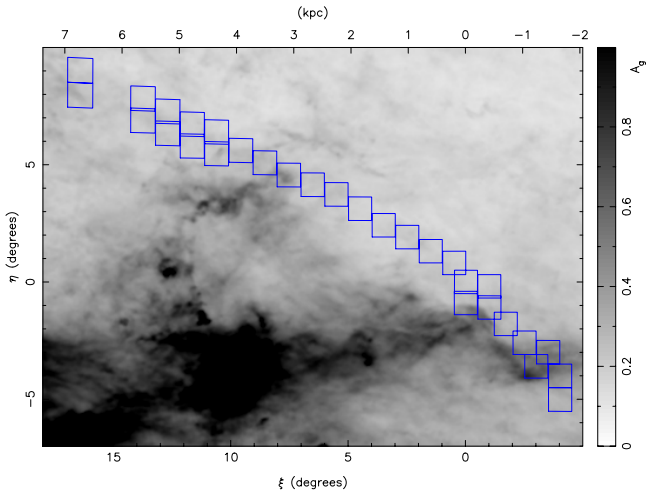}
}
\end{center}
\caption{Extinction in the g-band over the survey region, derived from the \citet{Schlegel:1998fw} maps. The extinction clearly varies considerably across this region of sky, but is not too extreme in the selected fields. The minimum and maximum values of $E({\rm B-V})$ within the CFHT fields (blue squares) are $0.04$~mag and $0.22$~mag, respectively, while the average and rms values are $0.07$~mag and $0.03$~mag, respectively.}
\label{fig:extinction}
\end{figure}

A further use of tidal streams stems from their susceptibility to small-scale variations in the gravitational potential. If $\Lambda$ Cold Dark Matter cosmology is correct, Milky Way-like galaxies should contain hundreds of dark matter sub-halos \citep{Klypin:1999ej,Diemand:2008hr} with masses similar to dwarf satellite galaxies. If these dark halo substructures actually exist in reality, they would change the galaxy from a smooth calm force-field into a ``choppy sea'' where streams and their progenitors are tossed hither and thither. The effect of this is that streams should become dynamically heated, with a significantly larger line of sight dispersion, velocity dispersion, and width \citep{2002MNRAS.332..915I,2002ApJ...570..656J}. These effects are even more striking in energy and angular momentum space and may soon be accessible thanks to the Gaia mission. It has also been shown that gaps in streams  provide a means to constrain the abundance of dark matter sub-halos \citep{2012ApJ...760...75C,2014ApJ...788..181N}.

While it had been suspected for many years that globular clusters dissolve in the tidal field of the Galaxy, the evidence for this process was hard to derive from the wide-field data (photographic plates) that were available before the advent of panoramic CCD surveys \citep{2000A&A...359..907L}.
Indeed, the stellar stream of the \Pal\ globular cluster, which is the highest contrast such structure currently known, was first detected only in the Sloan Digital Sky Survey (SDSS) commissioning data by \citet{2001ApJ...548L.165O}, who found evidence for a stellar tail extending out to $\sim 1\degg3$ from the cluster center. That initial detection was spatially limited by the sky coverage at the time, and subsequent SDSS data releases showed that the stream was substantially longer  \citep{Rockosi:2002gv,2003AJ....126.2385O,Grillmair:2006ih} than the first measurements suggested. From the SDSS Data Release 4 (DR4), \citet{Grillmair:2006ih} used a matched filter technique to construct a map that suggests that the stream spans a full $22\deg$ across the sky: we re-display this area of sky selected using a suitable color-magnitude box from DR10 in Figure~\ref{fig:photometric_fields}, where one can immediately perceive the presence of the stream stars. 

The tidal disruption of this system was carefully modeled by \citet{Dehnen:2004ez}, who showed that the globular cluster was in its very last stages of dissolution, the result of several intense shocks as it crossed the Galactic disk during its past evolution. Spectroscopic follow-up with the high-resolution spectrographs UVES and FLAMES on the Very Large Telescope \citep{Odenkirchen:2009js} indicated also that the stream possesses a moderate velocity gradient and very low velocity dispersion ($2.2\kms$). A recent wide-field spectroscopic survey with the Anglo-Australian Telescope has identified several other radial velocity members and extended the velocity gradient \citep{2015MNRAS.446.3297K}.

\begin{figure}
\begin{center}
\includegraphics[angle=0, viewport= 25 30 760 585, clip, width=\hsize]{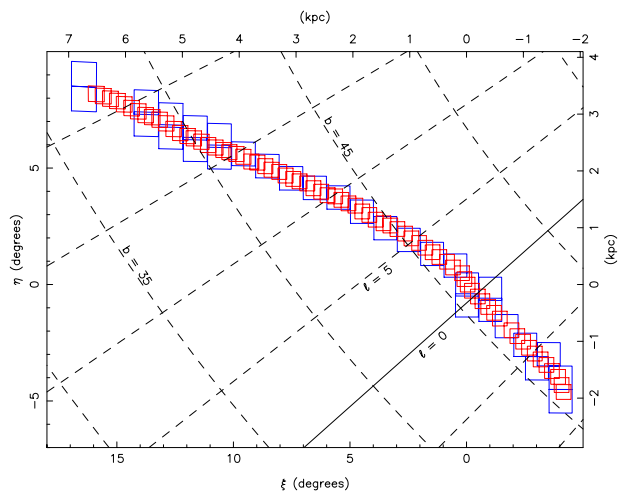}
\end{center}
\caption{Comparison of the CFHT (blue) and KPNO (red) survey fields. To help in understanding the orbit of this stream through the Milky Way, we have superposed a $5\deg\times5\deg$ grid in Galactic coordinates (dashed lines). Note that the $l=0\deg$ Galactic minor axis (which is shown with the continuous black line), passes very close to \Pal\, and is almost perpendicular to the stellar stream, implying that the structure is aligned roughly parallel to the Galactic disk, albeit at a high Galactic latitude of $45\deg$.}
\label{fig:survey_fields}
\end{figure}

The aim of the present series of papers is to explore the above ideas, using the stellar stream of the globular cluster \Pal\ both to constrain the global mass distribution in the dark halo, to probe dark matter substructure, and to investigate further the dynamical evolution of this particular low-mass stream. In this first contribution, we will present the analysis of the photometric surveys we have undertaken to map out and understand this structure.
In Section~\ref{sec:Photometric_Observations} we first describe the observations and data reduction. Section~\ref{sec:Gap_Analysis} presents the analysis for gaps and peaks along the stream; the analysis to constrain the distance is presented in Section~\ref{sec:distance} and a matched-filter analysis to map out the path, density and extent of the system is given in Section~\ref{sec:map}. Finally, in Section~\ref{sec:Discussion_Conclusions} we discuss the implications of these findings, and draw the conclusions from this study.

\begin{table}
\begin{center}
\caption{Properties of the globular cluster \Pal.}
\label{tab:properties}
\begin{tabular}{ccc}
\tableline\tableline
Parameter & value & source \\
\tableline
RA & $15^h 16^m 05^s.3$ & 1\\
Dec & $-00\deg 06' 41\sec0$ & 1 \\
$\ell$ & $0.8522$ & \\
$b$   & $+45.8599$ & \\
$E({\rm B-V})$ & $0.06$ mag & 2 \\
$(m-M)_0$ & $16.86$ & 3 \\
Distance & $23.5\kpc$ &  \\
Angular scale & $411\pc$ per degree & \\
${\rm [Fe/H]}$ & -1.3 & 4 \\
\tableline\tableline
\end{tabular}
\tablecomments{The sources are: 1 = \citet{DiCriscienzo:2006jv}, 2 = \citet{Schlegel:1998fw}, 3 = \citet{2011ApJ...738...74D}, 4 = \citet{2002AJ....123.1502S}. Rows without source information are derived from other table parameters.}
\end{center}
\end{table}

\section{Photometric Observations}
\label{sec:Photometric_Observations}

\subsection{CFHT data}
\label{sec:CFHT_data}

The primary source of photometry for this project was obtained with the MegaCam instrument at the 3.6m Canada France Hawaii Telescope (CFHT) during the 2006-2008 spring observing seasons. We mapped out the 30 ($1\deg\times1\deg$) fields shown with blue squares in Figure~\ref{fig:photometric_fields}, using ${\rm g}$ and ${\rm r}$ filters. The exposures were typically $3\times320~{\rm s}$ in $g$ and $3\times300~{\rm s}$ in ${\rm r}$. In addition, we obtained short $60~{\rm s}$ exposure fields offset by half a degree in Right Ascension to aid in the photometric calibration.

The processing of the CFHT data followed in an almost identical manner the procedure that was described in detail in \citet{2014ApJ...780..128I} (for ${\rm g}$- and ${\rm i}$-band observations of the halo of the Andromeda galaxy). Briefly, the data initially pre-processed by the CFHT `Elixir' pipeline \citep{2004PASP..116..449M}, were combined and measured with the Cambridge Astronomical Survey Unit (CASU) pipeline \citep{Irwin:2001eq}. The CASU software performs image detection, aperture photometry and measures a stellarity index, which is determined from the curve of growth. 
As in \citet{2014ApJ...780..128I}, a comparison of the CFHT MegaCam data with SDSS DR10 survey showed the presence of significant photometric differences that were spatially correlated with position on the MegaCam CCD detectors, with amplitude of up to $0.1$~mag peak-to-peak. Using bright stars with ${\rm 17<g<19}$ and ${\rm 17<r<18.5}$ in common between the SDSS and MegaCam surveys, we calculated a flat-fielding function to apply to each years' data. We used the following color equations to convert between SDSS and MegaCam filters:
\begin{displaymath}
     \begin{aligned}
{\rm g} &= {\rm g}_{\rm SDSS} - 0.185 ({\rm g}_{\rm SDSS} - {\rm r}_{\rm SDSS})\\
{\rm r} &= {\rm r}_{\rm SDSS} - 0.024 ({\rm g}_{\rm SDSS} - {\rm r}_{\rm SDSS})
     \end{aligned}
\end{displaymath}
which we found to give a better representation of the stars in our fields than the equations in \citet{Regnault:2009bk}. The global photometric calibration is achieved by solving for the unknown photometric zero-points of the CFHT fields, using the CCD overlaps (especially including those from the short exposure fields) and the SDSS which overlaps with a large number of our CFHT fields (as can be seen in Figure~\ref{fig:photometric_fields}). We iterated on this flat-fielding and zero-point determination to converge on a solution where the rms differences between the CFHT and SDSS photometry are $0.015$~mag and $0.013$~mag in the ${\rm g}$ and ${\rm r}$ bands, respectively. Henceforth ${\rm g}$ and ${\rm r}$ will refer to magnitudes on the CFHT system.

It is worth noting at this point that the foreground extinction in the region covered by the \Pal\ stellar stream is not negligible, and additionally it is fairly variable, as we show in Figure~\ref{fig:extinction}. Fortunately, however, the stream skirts the most extincted areas, giving rise to an average $E({\rm B-V})$ extinction over the CFHT fields of $0.07$~mag, with an rms of $0.03$~mag (the value at the cluster center is 0.06). This extinction is corrected for, assuming $A_{\rm g}/E({\rm B-V})=3.793$ and $A_{\rm r}/E({\rm B-V})=2.751$. (We assume ${\rm V = g - 0.59 (g-r) - 0.01}$, from \citealt{2005AJ....130..873J}). The CFHT catalog is listed in Table~\ref{tab:CFHT}.

\subsection{KPNO data}
\label{sec:KPNO_data}

We also secured photometric observations of the cluster and its stream using the Mosaic II imager on the Mayall 4m Telescope at Kitt Peak National Observatory (KPNO) on 2010, June 4--8. We defined 73 ($36\arcmin\times36\arcmin$) partially overlapping fields along the tidal stream, to complement the regions already observed with the CFHT. Our initial aim was to obtain Washington photometry to better discriminate dwarfs and giants along the stream, following the method of \citet{2000AJ....120.2550M}, and also to use deep U-band observations to constrain the metallicity of the stars as accomplished for main-sequence SDSS stars by \citet{Ivezic:2008fd}, which we hoped would also allow for better discrimination between stellar populations. During the first two nights of this KPNO run we successfully observed the fields shown in red in Figure~\ref{fig:survey_fields} in the DDO~51 and $M$-band (approximately $V$) filters, but unfortunately due to subsequent bad weather, the $C$-band (approximately $U$) and $I$-band observations we took were not of sufficient quality and spatial homogeneity to be useful to the present project. For brevity, we will use ${\rm D}$ to refer to magnitudes in the DDO~51 filter. The adopted exposure times for these fields were $1\times 100$~s in ${\rm M}$ and $1\times 500$~s in ${\rm D}$.

\begin{figure}
\begin{center}
\hbox{
\includegraphics[angle=0, viewport= 45 55 700 560, clip, width=\hsize]{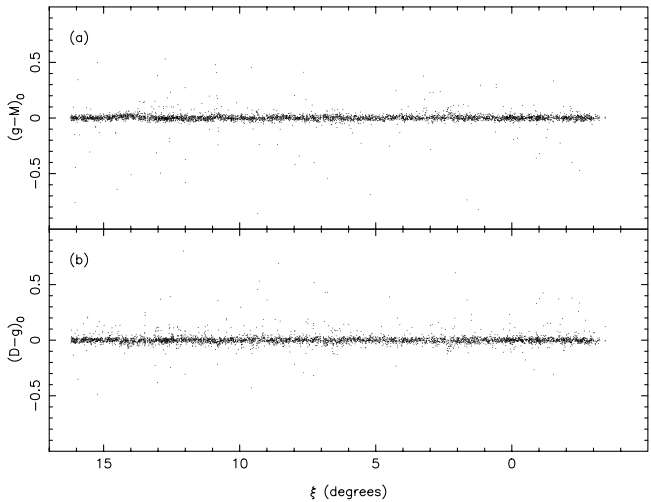}
}
\end{center}
\caption{Photometric differences between SDSS photometry of stars in the small color-magnitude window $0.3 < (g-r)_0<0.4$, $16<g_0<18$ and the DDO~51 and $M-$band magnitudes calibrated in this study. The rms of the differences (after clipping $5\sigma$ outliers) is $<0.03$~mags in each band.}
\label{fig:photo_offsets}
\end{figure}

\begin{figure*}
\begin{center}
\hbox{
\includegraphics[angle=0, viewport=  85 25 720 560, clip, height=7.8cm]{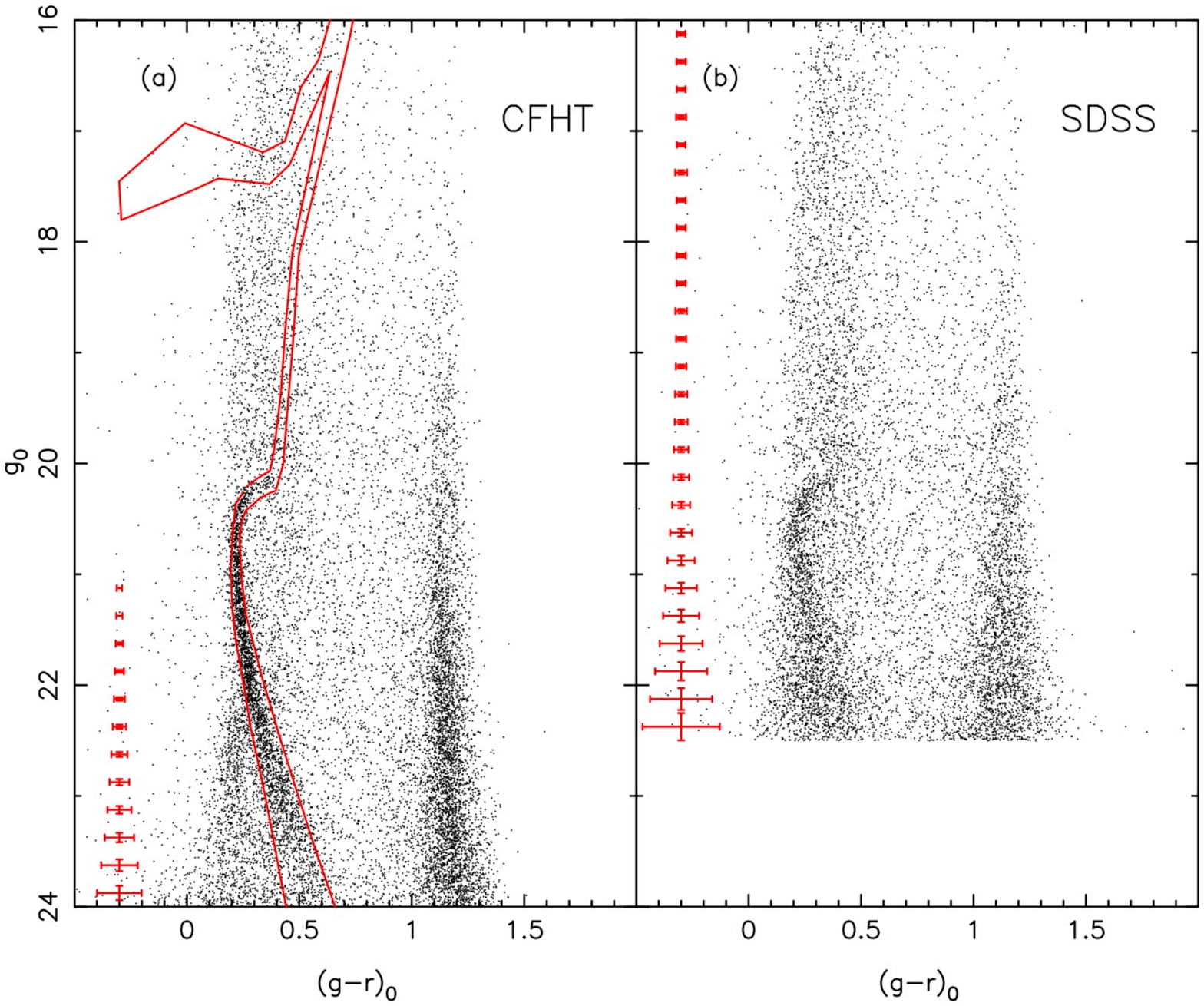}
\includegraphics[angle=0, viewport= 123 25 720 560, clip, height=7.8cm]{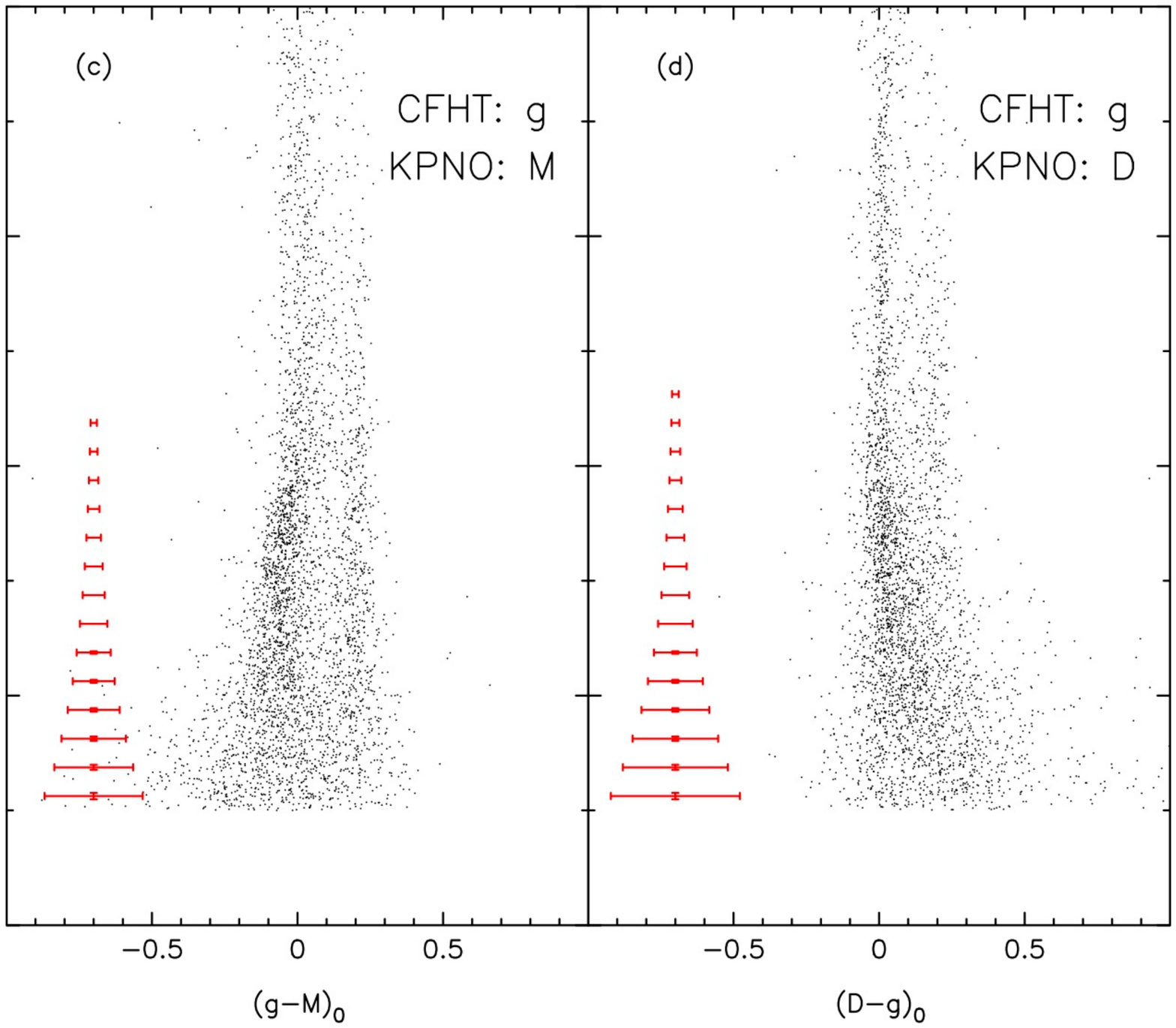}
}
\end{center}
\caption{Photometry of Palomar~5. Panels (a) and (b) compare the CFHT and SDSS photometry in a 1 square-degree field centered on the globular cluster remnant. The irregular polygon superimposed on (a) encompasses the stellar populations observed within the central $15\arcmin$.
The error bars show the average color and magnitude uncertainties for stars in the color range ${\rm 0.2<(g-r)_0<0.7}$, and are only shown when either the color or magnitude uncertainty exceeds 0.01~mag. Clearly the CFHT photometry provides a significant improvement over the SDSS in this region. In panels (c) and (d) we join the CFHT and KPNO photometry.}
\label{fig:Center_CMDs}
\end{figure*}

The Mosaic II images were processed using the CASU photometric pipeline, in a similar way to the CFHT photometry. However, these data suffered significantly from contamination by scattered light, which produces a ring-like additive contamination in the central CCDs of the mosaic. To correct for this, we used the IRAF algorithms {\it mscpupil} and {\it irmpupil} following the procedure outlined in \citet{1998ASPC..145...53V}. Unlike the case for the CFHT ${\rm g}$ and ${\rm r}$ bands, we lacked photometric reference data for the DDO~51 and ${\rm M}$-band photometry. Fortunately, the substantial overlaps allowed us to solve for the photometric differences, using fields taken during photometric conditions to anchor the offsets. We chose not to attempt to calculate the true photometric zero-points, leaving our DDO~51 and ${\rm M}$-band magnitudes on an instrumental system with arbitrary zero-point. In Figure~\ref{fig:photo_offsets} we have taken stars in the SDSS within a narrow color range ${\rm 0.3 < (g-r)_0<0.4}$ and with magnitudes ${\rm 16<g_0<18}$, and compared these with the KPNO photometry. Clearly the photometric differences are well-behaved over the spatial extent of the survey, indicating that the calibration of the KPNO photometry is uniform at the $\sim 0.03$~mag level. The resulting KPNO catalog is listed in Table~\ref{tab:KPNO}.

To give a first impression of the CFHT, SDSS and KPNO data and the photometric depth in each band, we display the CMDs of the central fields in Figure~\ref{fig:Center_CMDs}. The CFHT ${\rm g}$, ${\rm r}$ bands provide excellent data probing down to ${\rm g_0=24}$ with color uncertainty of $\delta({\rm g-r})_0=0.1$. The good precision around the cluster main sequence turnoff allows a much improved selection compared to the SDSS shown in panel (b). The KPNO data is of similar precision to the SDSS, however. 

Panel (a) of Figure~\ref{fig:Center_CMDs} shows a second main sequence structure approximately two magnitudes fainter than that of \Pal: this is due to the stream of the Sagittarius dwarf galaxy. Visual inspection of the CMDs in our fields shows that the Sagittarius stream is present in all of the fields at $\xi<5\deg$. This is a serious contaminating population whose effect has to be considered carefully when employing the CFHT photometry at $\xi<5\deg$ to its full depth; however, for ${\rm g_0<22}$ there is virtually no contribution from Sagittarius, and it should not pose a problem.

\begin{figure}
\begin{center}
\hbox{
\includegraphics[angle=0, viewport= 125 30 690 545, clip, width=\hsize]{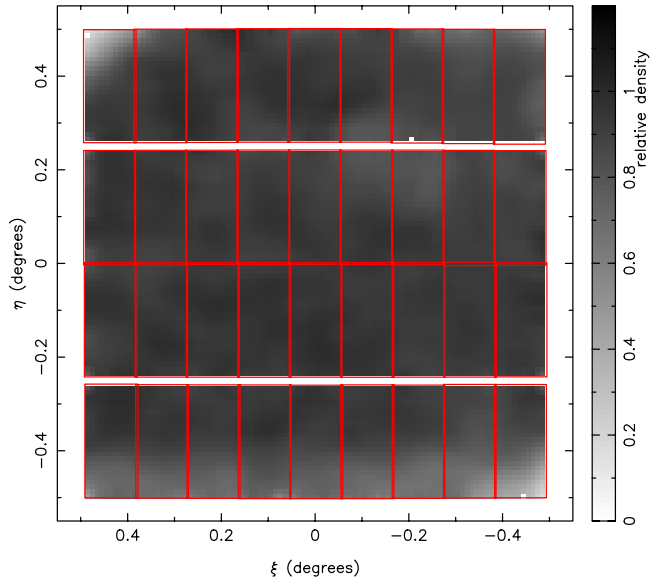}
}
\end{center}
\caption{Relative density of sources detected over the CFHT/MegaCam field of view. All stars with magnitudes ${\rm 20<g<23}$ that lie {\it outside} of the \Pal\ CMD selection box were used to construct this starcounts ``flat''. The tapering of the number of detected stars towards the edges of the instrument is primarily a result of the poorer image quality away from the optical axis of the camera.}
\label{fig:flat}
\end{figure}

\begin{figure*}
\begin{center}
\includegraphics[angle=0, viewport= 111 76 725 570, clip, width=\hsize]{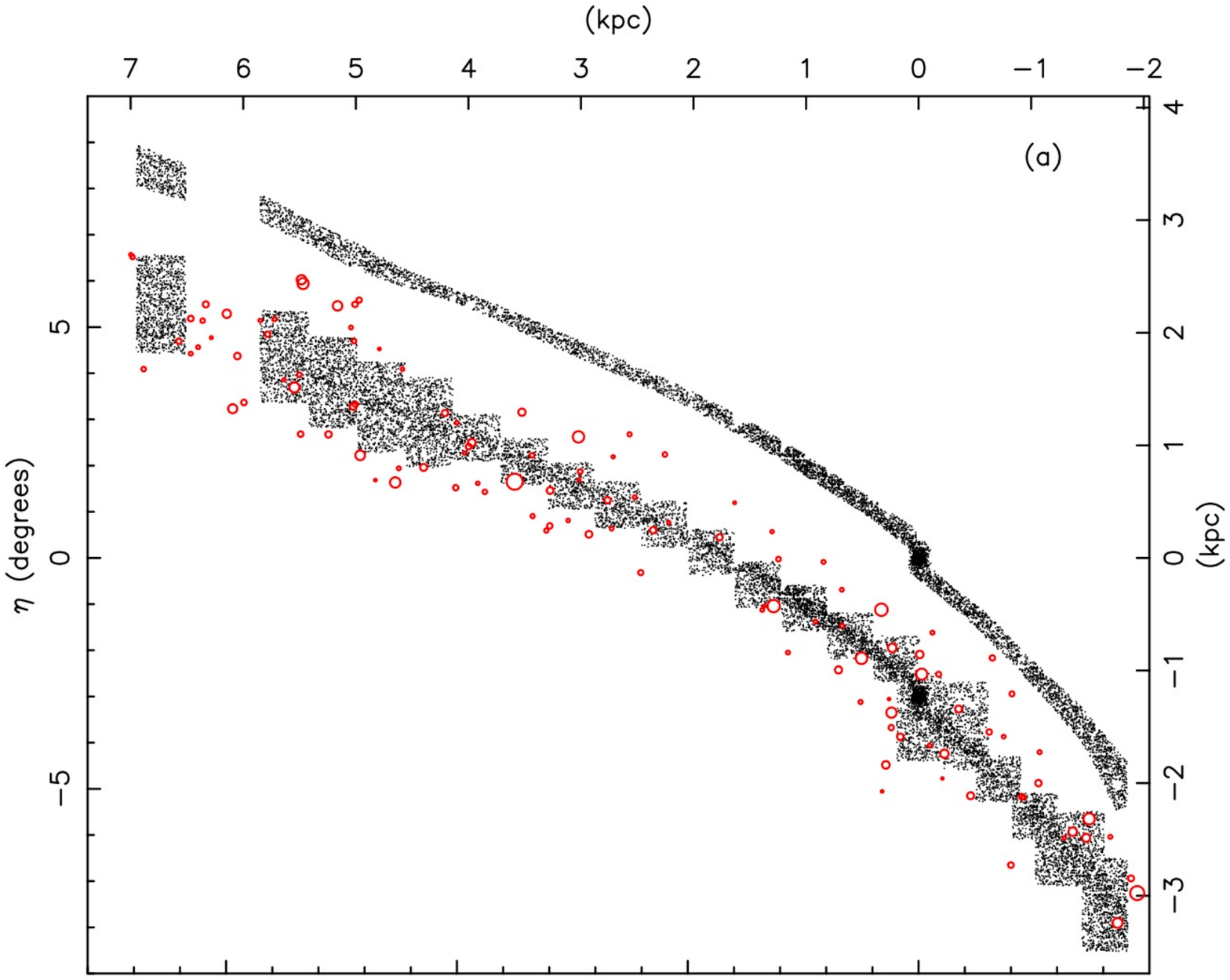}
\includegraphics[angle=0, viewport= 78 55 710 307, clip, width=\hsize]{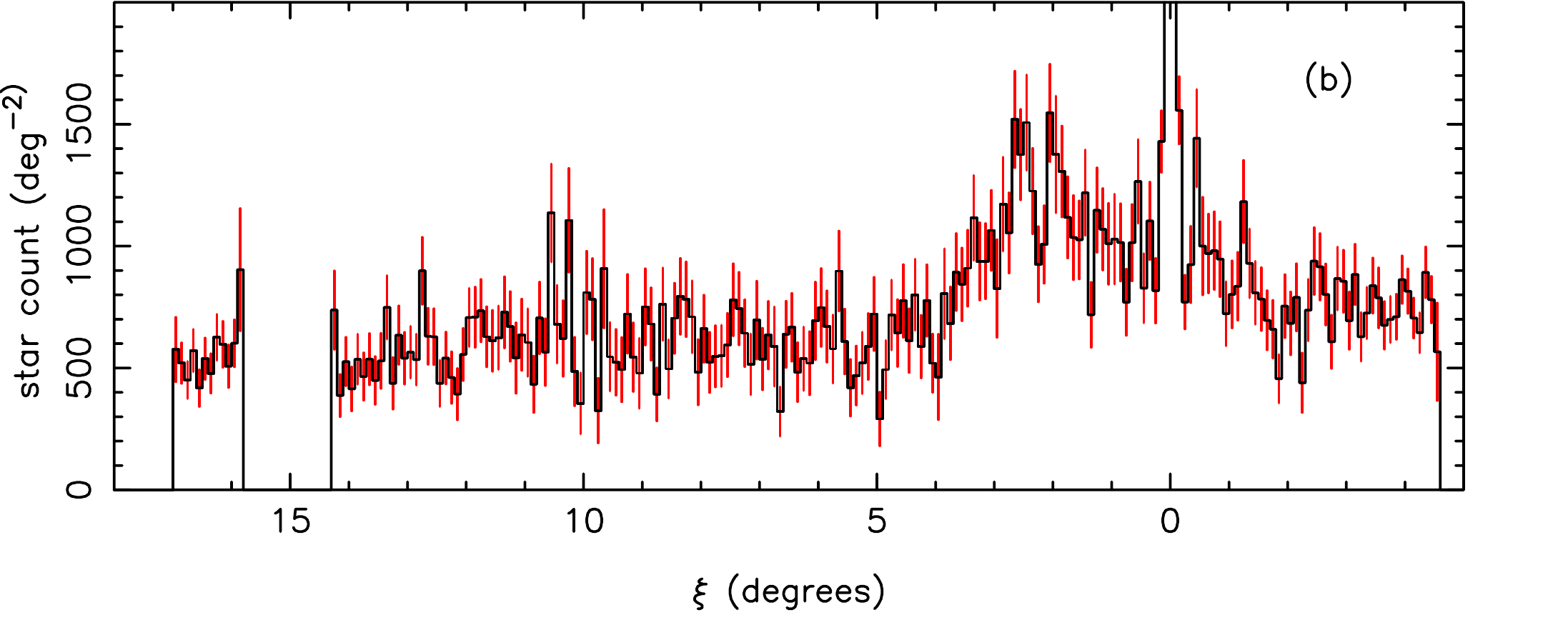}
\end{center}
\caption{The \Pal\ stellar stream derived from CFHT data. Panel (a) displays the CFHT stars, selected from the color-magnitude box surrounding the main-sequence between ${\rm 20<g_0<23}$ shown in Figure~\ref{fig:Center_CMDs}a. This sub-sample is shown shifted south by $3\deg$. A further sub-sample, spatially selected to follow the stream-like structure is shown at the actual ($\xi,\eta$) position. The red circles show the ($\xi,\eta$) positions of bright stars (also shifted South by $3\deg$); all survey stars within these circular regions are removed from the sample. The histogram in panel (b) displays the distribution of stellar density along the stream, properly flat-fielded, and limited to ${\rm 20 \le g_0 \le 23}$, to which magnitude the survey is homogenous.}
\label{fig:CFHT_stream}
\end{figure*}

\section{Gap and Peak Analysis}
\label{sec:Gap_Analysis}

Probably one of the most interesting questions we can attempt to answer with this dataset is whether there is evidence for gaps along this stellar stream that could be attributed to the gravitational disturbances induced by dark matter sub-halos. A first analysis therefore should be to count carefully the number of \Pal\ stars along the stream and determine whether their spatial distribution shows evidence for gaps that would not be expected from Poisson statistics.

Before presenting the CFHT star-counts map, it is important to appreciate the limitations of the camera. Away from the optical axis of Megacam, the image quality degrades significantly, which renders the star-galaxy discrimination more uncertain. As a consequence, the number of ``star'' detections changes over the field of view of the camera. The effect is shown in Figure~\ref{fig:flat}, where we show the number density of sources classified as stars within the magnitude range ${\rm 20<g_0<23}$ but that do not fall within the \Pal\ CMD selection box displayed in Figure~\ref{fig:Center_CMDs}a. It can be seen that the density of point-sources falls substantially towards the Northeast and Southwest corners of the mosaic. We use this detection density map as a ``classification flat-field'' to correct the observed number counts of \Pal\ stream stars, and to derive the corresponding uncertainties. In comparison, the spatial variations in completeness from field-to-field are relatively small, at the $\sim 2$\% level (see the results of the completeness analysis in Appendix A), and can safely be neglected.

Figure~\ref{fig:CFHT_stream}a shows the positions of the stars in the CFHT survey that have ${\rm 20<g_0<23}$ and lie within the CMD selection box (Figure~\ref{fig:Center_CMDs}a). The full survey region is displayed shifted $3\deg$ to the South, with the red circles marking the positions of very bright stars in the vicinity of the survey. Since the bright stars cause a local degradation in detection efficiency, any CFHT sources within these circular regions are removed from the catalog, and we correspondingly correct for the missing areas when counting the stars. 

A spatially-selected sub-sample is shown at the actual position of the stream. In panel (b) we display the corresponding star-counts, together with their $1\sigma$ uncertainties (as red bars), as a function of the standard coordinate $\xi$, in bins of size $0\degg1$.  
The completeness of this sample of stars (with ${\rm 20<g_0<23}$ that lie within the CMD selection box) is $> 80$\%.
It is worth noting that the uncertainties are due purely to Poisson noise, corrected by the starcounts flat-field. Over-densities and under-densities can be visually matched between the two panels. This star-counts profile contains an underlying ``background'' population of Galactic (and Sagittarius stream) stars that contaminate the \Pal\ sample. We will not remove them though at this juncture, as we are presently interested in placing an upper limit on the incidence of substructure, and obviously neither the Galaxy nor Sagittarius can be smoother than Poisson statistics allow. Their presence in the sample should not bias the result towards finding a smoother profile, barring the unlikely event that peaks in their spatial distribution fill in the gaps in the \Pal\ stream.

To detect the presence of possible gaps along the stream, we reproduce the analysis of \citet{2012ApJ...760...75C}, who used a gap-filter to search for holes in the \Pal\ stream with SDSS data. We implemented their ``$w_2$'' gap filter (which is their preferred filter of the two they examine, but which in any case gives very similar results to their ``$w_1$'' filter). The $w_2$ filter is defined as $w_2 (x) = (x^8 - 1) \exp (-0.559 x^4)$, which is designed to be $-1$ at $x=0$, crossing $0$ at $|x|=1$, rising to $\approx 1.61$ at $|x| \approx 1.40$, thereafter falling off rapidly to zero. The filter thus has a double-horned shape designed to be similar to the profiles of the sub-halo induced gaps seen in the simulations of \citet{2012ApJ...748...20C}. The characteristic width of the gaps selected by this filter is $\sim 2$, hence when using a kernel of 1 pixel, we should be particularly sensitive to gaps of characteristic size $0\degg2$. By selecting different kernel widths for this filter, one can search for gaps of different characteristic sizes. 

\begin{figure*}
\begin{center}
{\hbox{
\includegraphics[angle=0, viewport= 37 120 535 720, clip, width=8.8cm]{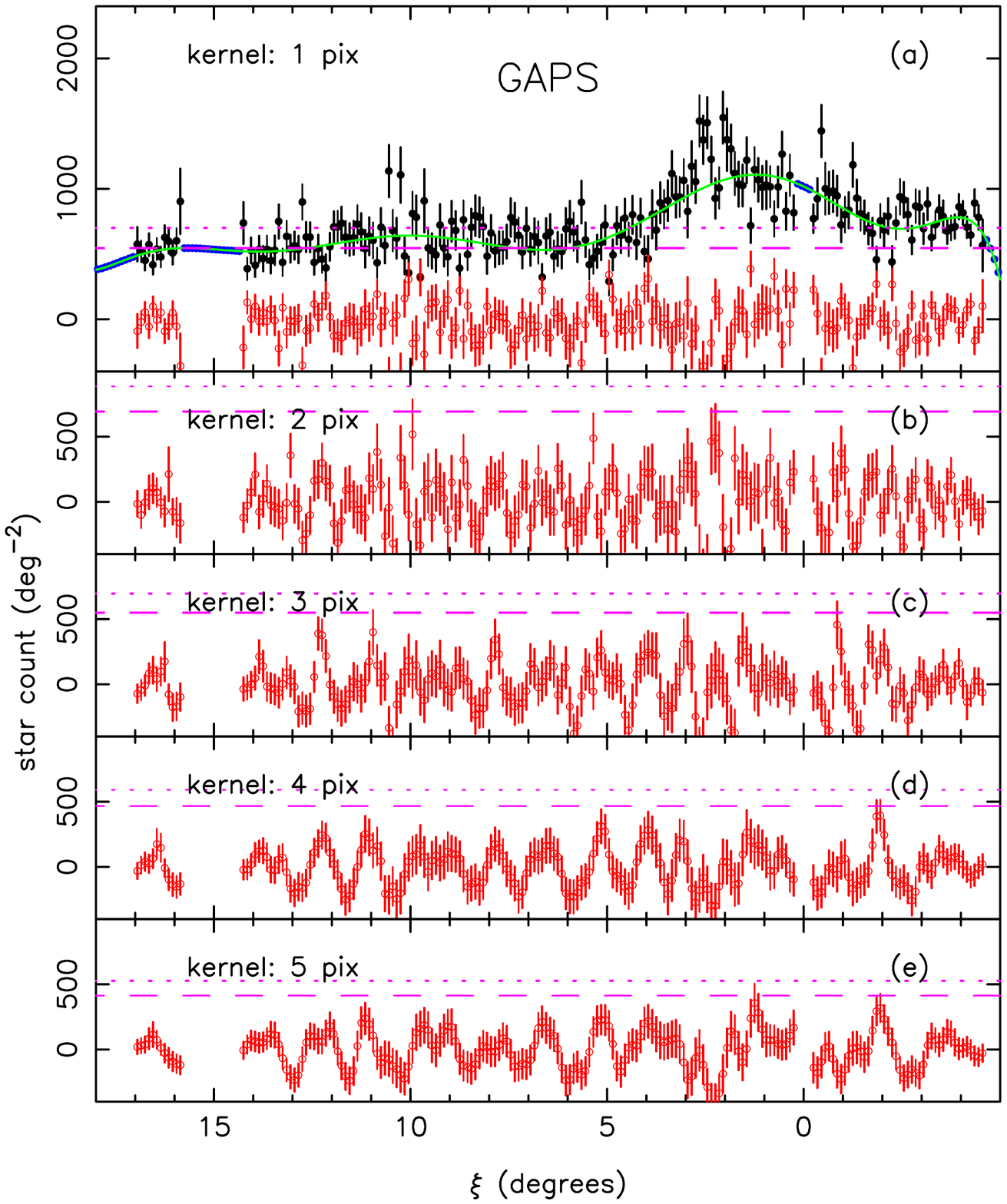}
\includegraphics[angle=0, viewport= 37 120 535 720, clip, width=8.8cm]{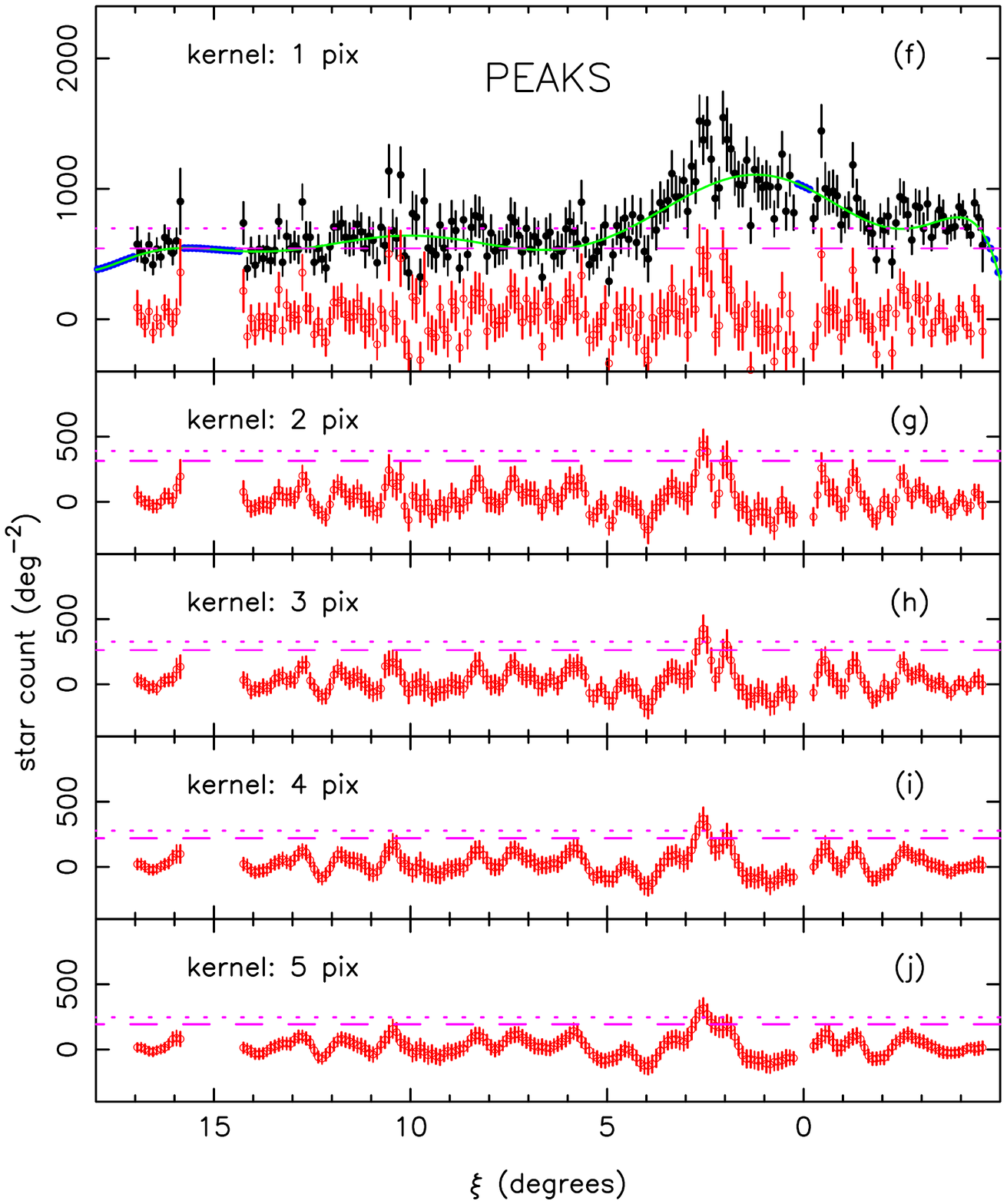}}}
\end{center}
\caption{Gap (left panels) and peak (right panels) analysis of the CFHT star-counts. The black dots in the upper panels reproduce the stream star-counts from Figure~\ref{fig:CFHT_stream}b, but with the counts in the central $0\degg5$ around the cluster removed. The green line shows a $9^{\rm th}$ order polynomial fit through the black dots. Where there is missing data, we fill in the profile with the value from the fitted polynomial (blue dots). Red dots show the result of applying the \citet{2012ApJ...760...75C} gap filter; from panel (a) to (e) the width of this filter kernel is increased from 1 to 5 pixels.
Positive spikes in the red distribution correspond to possible gaps in the stream. The pink dashed and dotted lines show, respectively, the level of the highest $10$\% and $1$\% spikes in Monte Carlo simulations drawing samples consistent with the polynomial fit. The observed gaps are therefore not highly significant. The right-hand panels show the same analysis, but for over-densities in the stream (using a peak-finding filter).}
\label{fig:gaps}
\end{figure*}

We note that \citet{2012ApJ...760...75C} chose to split their SDSS catalog into bins in linear increments of $0\degg1$, which motivated our bin size choice in Figure~\ref{fig:CFHT_stream}b.

In Figure~\ref{fig:gaps}a--e we display our implementation of this method for searching for gaps in the stream. The black points and error bars in panel (a) recall the profile previously shown in Figure~\ref{fig:CFHT_stream}b, but with the data in the central $0\degg5$ region around the cluster excised. We fit a $9^{\rm th}$ order polynomial to these data (taking in account the uncertainties), which is shown as a green line. At the locations where there is missing data, we filled in the distribution using this polynomial (blue points).

The red points in panel (a) display the results of applying the \citet{2012ApJ...760...75C} $w_2$ filter with a kernel width of one ($0\degg1$) pixel to this profile. Convolution with the gap filter causes gaps in the data to be visible as peaks in the filtered profile, which can be readily verified by visual inspection of the diagram. To obtain a first estimate of the expected noise in the filtered gap profile, we resampled the input profile 1000 times drawing from the black data points (consistent with their associated uncertainties); the resulting scatter is shown with the red error bars. One appreciates from this that many of the gap detections lie consistently above the average level of the red profile.
However, we then attempted to quantify what the expected level of the gap detections should be if the profile were smooth. To this end we used the $9^{\rm th}$ order polynomial fit (green line) together with the uncertainties associated with the real profile (black) to generate 10000 realizations of that smooth profile. For each realization we identified the most prominent gap detection. The pink dashed and dotted lines in panel (a), show respectively, the level of the 10\% and 1\% most prominent gaps in this Monte-Carlo experiment. For the one pixel kernel, the most prominent of the observed gap detections lie at about the 10\% level. Based on these data we deduce that there are no significant gaps of typical size $\sim 0\degg2$.

Panels (b) to (e) show the result of increasing the kernel size in $0\degg1$ increments up to $0\degg5$. Similar to the finding for panel (a), no significant gap is found for any of these kernel sizes.

\begin{figure*}
\begin{center}
\includegraphics[angle=0, viewport= 15 55 750 300, clip, width=\hsize]{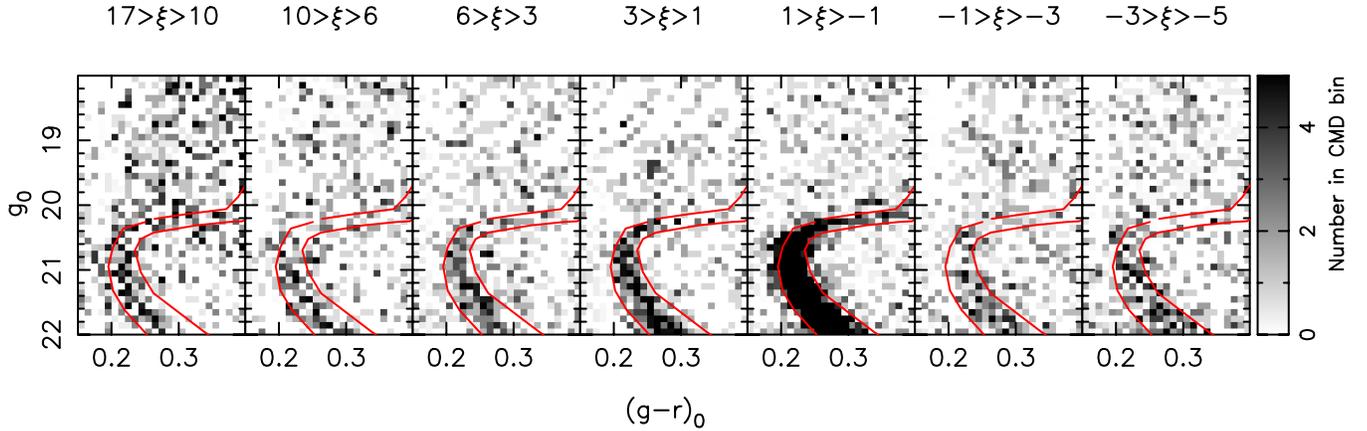}
\end{center}
\caption{Background-subtracted Hess diagrams of seven regions along the \Pal\ stream. The spatial interval in the standard coordinate $\xi$ is marked over each panel. The red lines (a portion of the CMD selection box shown previously in Figure~\ref{fig:Center_CMDs}) encompass the main-sequence turnoff region of the cluster,  as is very clear in the fifth panel. However, this population is also present in all of the stream regions, albeit at slightly fainter magnitudes at positive $\xi$, and slightly brighter magnitudes at negative $\xi$.}
\label{fig:Hess}
\end{figure*}

Another interesting question is whether there is evidence for peaks in the profile. To address this issue, we repeated the above analysis using the following peak filter: $w_{peak}(x) = {{\log[16]}\over{\Gamma[5/4]}} \exp (- 16 \log[2] x^4)$, which has full-width at half-maximum of unity, and integrates to 1. This filter approximates a (smoothed) top-hat, appropriate for detecting density peaks; we also chose it for its functional similarity to the $w_2$ gap filter discussed above. The results for kernels of width 1 to 5 pixels are shown in panels (f)--(j). It is evident that for kernel widths exceeding 3 pixels ($0\degg3$), the observed peak at $\xi=2\degg5$ is more significant than 99\% of random realizations drawn from the smooth polynomial fit. No other locations along the stream stand out, however.

\section{Distance variation}
\label{sec:distance}

The \Pal\ stream extends over an extremely large distance perpendicular to the line of sight, approximately $10\kpc$ (note the scale on the upper and right-hand axes of Figure~\ref{fig:CFHT_stream}a). At the same time, it is only $23.5\kpc$ distant from us \citep{2011ApJ...738...74D}, so we could expect to see substantial variations in Heliocentric distance from one end of the structure to the other. The distance gradient is also a key ingredient in any dynamical modeling, and should prove useful as a constraint in the modeling of the Galactic mass distribution.

\begin{figure*}
\begin{center}
\includegraphics[angle=0, viewport= 15 55 750 330, clip, width=\hsize]{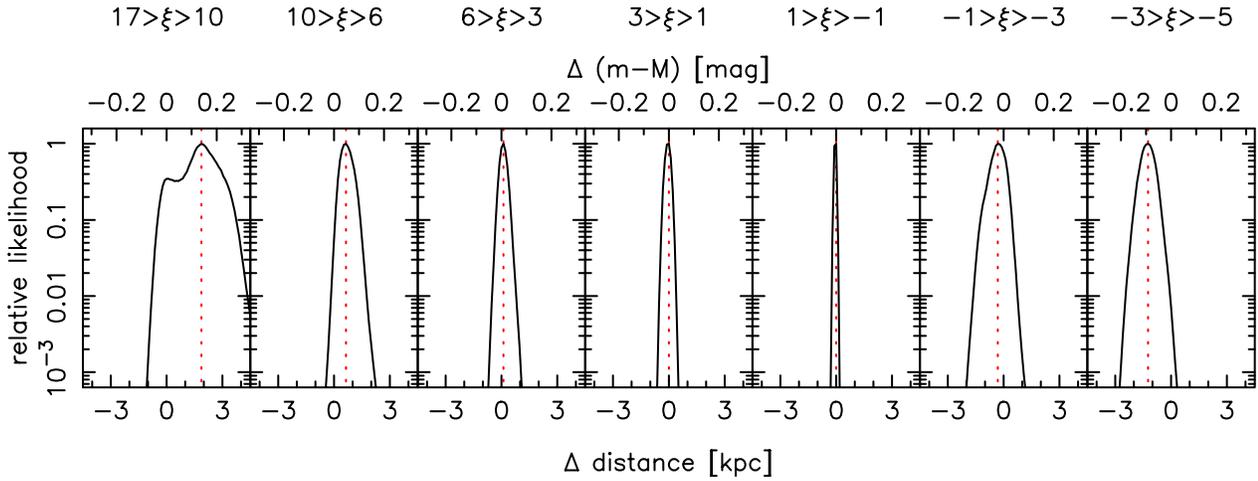}
\end{center}
\caption{Relative likelihood of the distance to the stream population in the same spatial regions as in Figure~\ref{fig:Hess}. The scale at the top of the panels is shown in magnitudes, on the bottom in $\kpc$.}
\label{fig:distance}
\end{figure*}

\begin{figure*}
\begin{center}
\includegraphics[angle=0, viewport= 25 80 760 585, clip, width=\hsize]{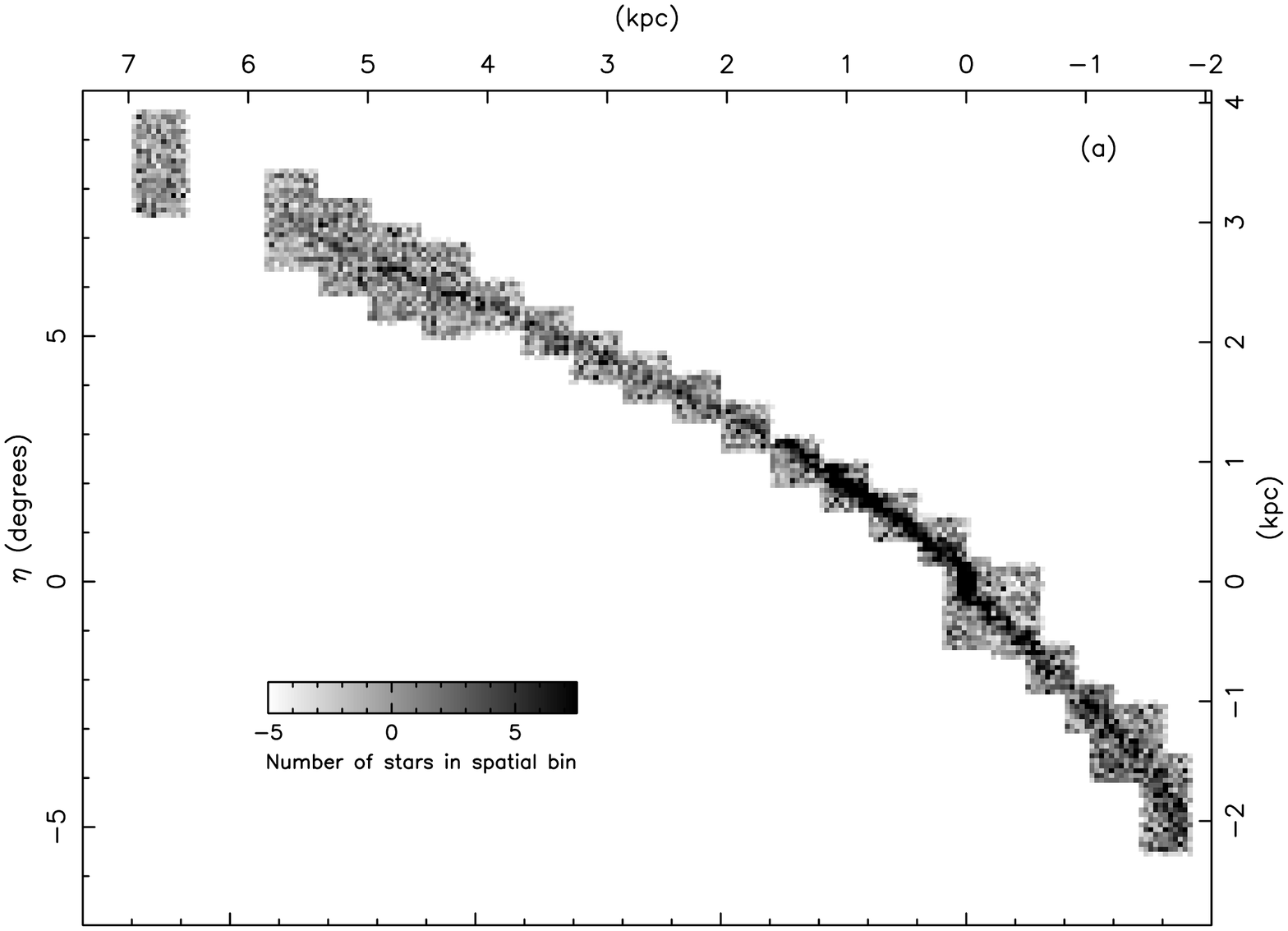}
\includegraphics[angle=0, viewport= 25 30 760 214, clip, width=\hsize]{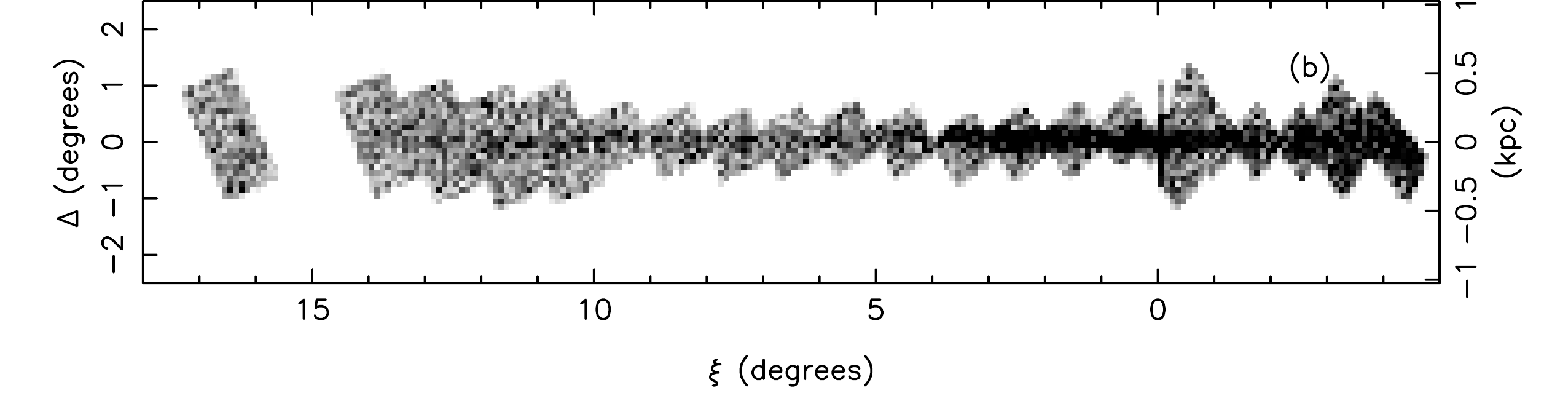}
\end{center}
\caption{(a): Matched filter map of the \Pal\ stream, using the CFHT data down to $g_0=24$. The insert shows the density of stars per $0\degg1 \times 0\degg1$ spatial bin. The CMD filter has been shifted slightly in magnitude by $0.009 \big[ {{\xi}\over{\rm 1 degree}} \big]$ to account for the distance gradient revealed in Figure~\ref{fig:distance}. The scatter in the background off-stream regions in this map is approximately 2.4 counts. Panel (b) shows a re-projection of these data binned along the $\xi$ coordinate axis, oriented so that $\Delta$ measures distance perpendicular to our stream fit.}
\label{fig:matched_filter}
\end{figure*}

Inspection of Figure~\ref{fig:Center_CMDs}a shows that the CMD region between ${\rm 20<g_0<22}$ is particularly useful for the task of measuring relative distances, because the data there are of excellent photometric accuracy, the sequence is clearly very tight, and has a hook-like shape that is sensitive to offsets in magnitude. We further chose to impose a color cut ${\rm 0.15 < (g-r)_0<0.4}$ for simplicity, although this choice has virtually no consequences on our distance measurements.

Away from the cluster centre, the foreground (and background) Milky Way stars dominate the color-magnitude diagram, so we need a means to subtract or correct for these contaminating populations in order to measure the stream stars of interest. One way would have beed to construct a contamination model that depends on CMD position and spatial position, in a similar way to \citet{2013ApJ...776...80M}. However, the relatively small spatial extent of the survey perpendicular to the stream makes such an undertaking rather difficult. Instead, we build a model based on the outermost regions of the survey at $\xi>5\deg$ that also lie outside of the stream spatial selection region shown in Figure~\ref{fig:CFHT_stream}. The region of the Milky Way that the stream covers can be seen in Galactic coordinates in Figure~\ref{fig:survey_fields}; note that the range of Galactic latitude is relatively modest, so the contaminating populations should not vary substantially. 

We will assume that the ratio of brighter stars in the range ${\rm 18<g_0<19.5}$ (selection $A$, say) to fainter stars in the range ${\rm 19.5<g_0<22.0}$ (selection $B$) remains constant (with both samples curtailed to ${\rm 0.15 < (g-r)_0<0.4}$). The validity of this assumption can be tested by comparison to the SDSS. To this end we select SDSS stars adjacent to the stream within the range $0\degg5 < |\Delta| < 2\deg$ (the coordinate $\Delta$ is defined in Figure~\ref{fig:matched_filter})\footnote{We also reject stars within $1\deg$ of the globular cluster M5.}, and count the ratio $N_A/N_B$ of the number of stars in the brighter to the fainter box in the seven spatial regions along the stream listed in Table~\ref{tab:distances}. Averaged over all the spatial regions, we find $N_A/N_B=0.598$, and all of the seven spatial regions are consistent with this value to within one standard deviation of Poisson counting uncertainties. We also find no significant gradient in $N_A/N_B$ along $\xi$. Having validated our assumption that $N_A/N_B$ is constant, we can now use it to estimate the contamination within any selected spatial region in the CFHT survey, simply by measuring the corresponding number of Milky Way stars within ${\rm 18<g_0<19.5}$ and ${\rm 0.15 < (g-r)_0<0.4}$.

Figure~\ref{fig:Hess} shows the color-magnitude distribution of stars in the seven different locations along the stream. The spatial selection of Figure~\ref{fig:CFHT_stream} has been applied, and the contaminating population removed, using the method described above. To aid visual  comparison, we have overlaid with red lines in each panel of Figure~\ref{fig:Hess} the CMD selection box of Figure~\ref{fig:Center_CMDs}a. The Hess diagram for the $1>\xi>-1$ sample shows the strong signal from the cluster main-sequence turnoff.  However, one can also perceive the same population in all other panels, at approximately the same distance modulus as the cluster population. More careful inspection shows that the \Pal\ populations at negative $\xi$ lie towards the brighter edge of the red lines, while those at positive $\xi$ lie at slightly fainter magnitudes, implying a relatively modest distance variation over the structure.

To quantify the distance shifts, we fitted the \Pal\ population in each one of the spatially-selected samples. The CMD structure of the background population, $B_{\rm g-r, g}$ (a function of ${\rm g-r}$ color and magnitude ${\rm g}$), was modelled in the same way as discussed above. To construct a model of the target \Pal\ population of interest $P_{\rm g-r, g}$, we took the photometry within the central $15\arcmin$ of the cluster and binned these into an array of ${\rm 0.01~mag \times 0.01~mag}$ intervals in color and magnitude. This allows us to calculate the likelihood that the \Pal\ population in the selected region has a offset in distance modulus of ${\rm \Delta(M-m)}$, as follows:
\begin{equation}
\ln \mathcal{L}[{\rm \Delta(M-m)}]= \sum_{\rm data} \ln ( F P_{\rm g-r, g+\Delta(M-m)} + B_{\rm g-r, g} ) \, ,
\end{equation}
where $F$ is a multiplicative factor to ensure that the sum of the background model plus the \Pal\ model give the total number of stars detected in the spatial region. The peak value of $\ln \mathcal{L}[{\rm \Delta(M-m)]}$ yields the maximum likelihood estimator of the line of sight distance offset of the population from that of the cluster center.

The likelihood of the combined background$+$stream model given the data in each spatial region is shown in Figure~\ref{fig:distance}. The most likely values of the distance of the stream stars (listed in Table~\ref{tab:distances}) are close to that of the cluster, and rule out a substantial curvature to the orbit. The relation:
\begin{equation}
\Delta {\rm (m-M)}(\xi) = 0.00890 \xi - 0.00186 \xi^2 + 0.00016 \xi^3 \, ,
\end{equation}
(with $\xi$ in degrees) provides a simple cubic model fit to these magnitude offsets.

\begin{table}
\begin{center}
\caption{Most likely value of relative distance along the stream. The uncertainties correspond to $68.3$\% confidence intervals.}
\label{tab:distances}
\begin{tabular}{crr}
\tableline\tableline
Selection & $\Delta$ distance [$\kpc$] & $\Delta {\rm (m-M)}$ [mag]  \\
\tableline
\hskip 0.12cm $17\deg > \xi > 10\deg$ &   ~$1.87\pm1.21$ &   ~$0.14\pm0.09$ \\
\hskip 0.00cm $10\deg > \xi > 6\deg$  &   ~$0.64\pm0.40$ &   ~$0.05\pm0.03$ \\
\hskip 0.10cm $6\deg > \xi > 3\deg$    &  ~$0.11\pm0.27$ &   ~$0.01\pm0.02$ \\
\hskip 0.10cm $3\deg > \xi > 1\deg$    &  $0.00\pm0.27$ &  $0.00\pm0.02$ \\
\hskip 0.30cm $1\deg > \xi > -1\deg$  &   $0.00\pm0.13$ &  $0.00\pm0.01$ \\
\hskip 0.10cm $-1\deg > \xi > -3\deg$ &   $-0.32\pm0.54$ &  $-0.02\pm0.04$ \\
\hskip 0.10cm $-3\deg > \xi > -5\deg$  &  $-1.24\pm0.54$ &  $-0.09 \pm0.04$\\
\tableline\tableline
\end{tabular}
\end{center}
\end{table}

\section{Matched Filter Map}
\label{sec:map}

Given that the CFHT photometry is significantly more accurate and deeper than the SDSS, it is well worth using it to present an updated map of the system. Following previous studies  \citep{Rockosi:2002gv,Grillmair:2006ih}, we decided to employ the matched-filter technique to better bring out the population of interest. This method uses a signal template (the cluster CMD) and a noise template (the contamination CMD), to construct a weighting filter that maximizes the signal-to-noise of the output map. We implemented a slight modification,  shifting the template CMD to account for the distance gradient measured in Section~\ref{sec:distance}. A linear shift of $0.009$~mag per degree in $\xi$ was adopted (derived from a simple linear fit to the data in Table~\ref{tab:distances}).

Figure~\ref{fig:matched_filter} shows the resulting map, where we have used all of the data down to a limiting magnitude of ${\rm g_0=24}$. The inserted grayscale wedge shows the level of these counts. 
The pixel to pixel r.m.s. scatter in regions outside of the stream spatial selection box at $\xi>5\deg$ is 2.4 counts, implying that the visually-obvious structure that is seen from $\xi=-5\deg$ to $\xi=15\deg$ can be confidently considered to be real.

We fitted the following cubic polynomials to this map, to trace the location of the two arms. The trailing arm fit is:
\begin{equation}
\eta_{\rm trailing}(\xi) = 0.211 + 0.768 \xi - 0.0305 \xi^2 + 0.000845 \xi^3 \, ,
\end{equation}
while the leading arm is modeled as:
\begin{equation}
\eta_{\rm leading}(\xi) =-0.199 + 0.919 \xi + 0.0226 \xi^2 + 0.0123 \xi^3 \, ,
\end{equation}
where the $\xi$ and $\eta$ coordinates are in degrees.
These models are used to calculate the perpendicular distance of the stars in the survey to the two stream arms, and ``straighten-out'' the structure, as displayed in panel (b). Obviously, this procedure gives rise to artifacts near the cluster itself, but it also highlights the fact that the stream is very thin, especially at large distance in the trailing arm.

\begin{figure}
\begin{center}
\includegraphics[angle=0, viewport= 32 32 478 478, clip, width=\hsize]{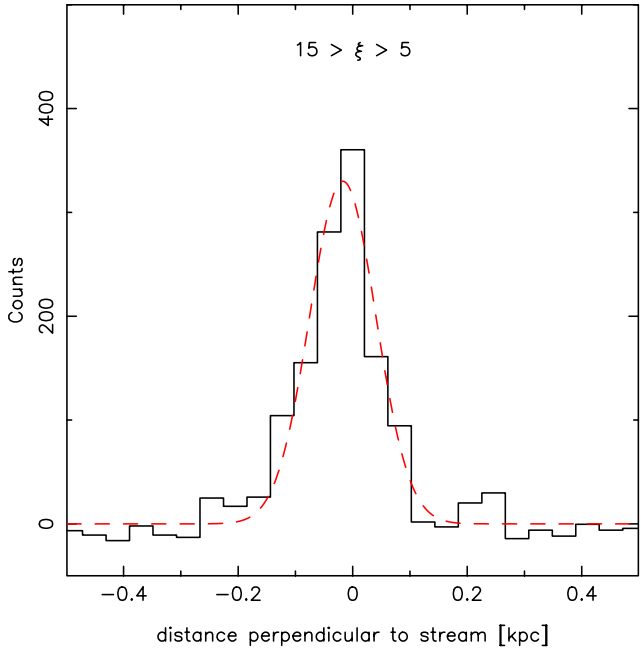}
\end{center}
\caption{The distribution of stars perpendicular to the stream, measured in the interval $15\deg >\xi>5\deg$ from our matched filter map (Figure~\ref{fig:matched_filter}). The dispersion of the fitted Gaussian (red-dashed line) is narrow: only $58\pc$.}
\label{fig:width}
\end{figure}

To quantify this further, in Figure~\ref{fig:width} we co-added the matched filter map between $15\deg > \xi > 5\deg$. The red dashed line shows a Gaussian fit to this width distribution, which has a dispersion of only $58\pc$.

\section{Discussion and Conclusions}
\label{sec:Discussion_Conclusions}

The gap analysis presented in Section~\ref{sec:Gap_Analysis} shows that our deep CFHT data do not support the presence of significant gaps along the stream, of characteristic width up to $1\deg$ ($\sim 400\pc$). The origin of the difference between our results and those of \citet{2012ApJ...760...75C} is difficult to establish, but it is likely that it is due to variations in homogeneity of the SDSS as one approaches the limiting magnitude of that survey. Furthermore, \citet{2012ApJ...760...75C} applied their analysis on a  matched-filter map, which obviously amplifies the contribution of certain stars. Those highly-weighted stars could well have color and magnitude values where the SDSS is less homogeneous. This latter concern was the main reason we chose to search for gaps in pure stars-counts rather than in our matched filter map. 

Turning our attention to peaks instead of gaps, our data show that the stream profile is remarkably smooth, with only a single significant overdensity, at $\xi=2\degg5$. This feature can be seen directly in the star map of Figure~\ref{fig:CFHT_stream}. Considerations based on epicycle theory suggest that there should be many overdensities along long streams from low-mass progenitors, at the points where stars slow down and bunch together in their migration outwards from the cluster (see, e.g., \citealt{2012MNRAS.420.2700K}). The overdensities seen in matched filter maps of \Pal\ from SDSS data appear consistent with these predictions \citep{2015ApJ...803...80K}. However, we corroborate only a single one of the \citet{2015ApJ...803...80K} overdensities: that at $l \cos(b) = 3\degg04$ (see their Figure~3), which they list as a $10.31\sigma$ detection. Their second most significant peak (away from the immediate vicinity of the cluster) lies at $l \cos(b) = 4\degg14$, and has significance of $7.43\sigma$; our profile from deeper data, with higher photometric accuracy, shows no hint of a peak at that location. We suspect that the cause of these differences is again the amplification of SDSS inhomogeneities by the particular matched filter that was used to construct the stream map.

One of the major differences between the measurement of peaks and gaps in earlier SDSS work and the present analysis is in the treatment of the contaminating populations due to stars from the foreground Milky Way, and from the (background) Sagittarius stream. In earlier work, matched-filter maps were first constructed to better isolate the population of interest, and then the contaminating populations were subtracted off, for instance (\citealt{2015ApJ...803...80K}) by using a heavily-smoothed version of the matched filter map as a background estimator. Note however, that when the contaminating population is subtracted off, the unavoidable subtraction errors may masquerade as gaps or peaks in the stellar stream (for instance, if the contaminating population possesses small-scale spatial structure in reality, subtracting off a smoothed version of it will produce fake holes or lumps in the population of interest).

In the analysis presented in Section~\ref{sec:Gap_Analysis} the contamination was not explicitly subtracted. We believe that this approach is the most conservative in the present context, since the presence of contaminating populations can only render the combined profiles less smooth. Our approach instead was to apply gap and peak-finding filters directly to the raw star-counts profiles, since these filters naturally search for localized gaps or enhancements. 

We could of course have implemented our gap and peak search analysis after subtracting out an estimate for the contamination. Note that if the contamination were a simple constant level along the profile, the results would have been precisely identical to what we obtained in Section 3. Since the peak and gap filters implement a local measurement, as long as the contamination varies slowly with position, our gap/peak analysis on the contamination-subtracted profile would yield very similar results to what we already obtained. The contamination should only affect our conclusions if it varies on a similar spatial scale as the substructure in the stream population. Nevertheless if this were the case, to produce the result that we found, the contamination would have to be perverse in the sense that it would have to have peaks where the stream has significant gaps, and gaps where the stream has significant peaks. This seems extremely implausible.

Hence, given that we found the profile of stream plus contamination to be smooth, we deliberately decided not to repeat the gap/peak analysis on the contamination-subtracted profile, as it would have weakened our result by potentially adding background subtraction errors to the analysis. So while we cannot rule out that the contaminating populations affect our gap/peak analysis, they do not hinder our ability to place an upper limit to the deviations from smoothness of the stellar stream. Note also that most observational errors (e.g., flat-fielding, field-to-field variable depth, field-to-field variable image quality, photometric zero-point errors, classification errors), will again have the tendency to create profile artifacts that appear as peaks or gaps.

We were initially struck by the apparent periodicity of the gaps and peaks in the lower panels of Figure~\ref{fig:gaps}. However, the Fourier Transforms of these profiles show no significant signal, and other tests we made, for instance, trying to fit sinusoids to the profiles, showed that any coherence does not persist over more than a few degrees. Given that the CFHT Megacam camera is $1\deg$ wide, and has a significant center-to-edge variation in sensitivity, there is good reason to be very skeptical about the reality of this apparent periodic signal. It will be interesting to revisit this issue with deep photometry from a different instrument.

The next steps to interpret these new CFHT findings will require extensive numerical modeling of the stream, to ascertain the extent to which the observed smoothness can place limits on the presence of dark matter sub-halos. Disentangling the effects from the ``normal'' secular evolution of the cluster will be important, and it is possible that the low incidence of peaks may also help to establish the dynamical history of this intriguing object.

A feature which appears particularly promising to us is the finding that the stream is very thin in the distant region $15\deg>\xi>5\deg$, with a typical width of only $58\pc$. These far-flung stars are among the first that were lost from the cluster and have had many Gyrs to probe the spatial variations in the Galactic potential. We suspect that the extremely thin morphology of the stream found here already places stringent conditions on the Galactic potential, strengthening the arguments made in \citet{2015ApJ...799...28P}.

We have seen in Section~\ref{sec:distance} that the cluster's stellar populations are clearly present over the majority of the CFHT survey region, confirming the $22\deg$ extent claimed by \citet{Grillmair:2006ih}, although the cluster most probably extends further to the West, out of the SDSS and CFHT survey regions. The good photometric precision within a magnitude of the main sequence turnoff also allowed us to place strong constraints on the relative line-of-sight distances to the stream along its length. The observed variations (Figure~\ref{fig:distance}), turned out to be very modest, deviating from equidistance by only $2\kpc$, and then only in the outermost regions. The direction of the measured gradient, as well as the magnitude of the distance shift, is in good agreement with the N-body simulations of \citet{Dehnen:2004ez}. In a future contribution we will use these constraints along with the positional and kinematic information on this system to model the disruption of the cluster and determine the structure of the Galactic gravitational potential.

Finally, we note that the matched-filter map we have constructed (Figure~\ref{fig:matched_filter}) can be used to weed out possible interlopers in kinematic surveys of the system (e.g. \citealt{Odenkirchen:2009js,2015MNRAS.446.3297K}). Likewise, the DDO~51 photometry, which we have briefly presented here, will be used in the next contribution to aid in decontaminating a new spectroscopic sample.

The \Pal\ stream thus appears to be very fine and very long, with a delicate structure that may well provide us with uniquely powerful information on both the global distribution of the dark matter in our Galaxy, and its ability to cluster on small-scales.

\bibliography{ms}
\bibliographystyle{apj}

\eject

\section*{Appendix A: Completeness Analysis}
\label{sec:Completeness_Analysis}

To examine the completeness as a function of magnitude and spatial location in the survey, we undertook a completeness analysis of our CFHT photometry. To this end, we re-reduced our frames using the point-spread function (PSF) package DAOPHOT \citep{Stetson:1987fx}. On each CCD of each frame we added 1000 artificial stars at random positions and with random magnitudes (within a ten-magnitude range that extended to approximately two magnitudes below the faint limit of the data). This number of artificial stars per frame is sufficiently low so as not to affect substantially the crowing of the fields. The frames were then re-processed with the CASU software in an identical manner to the unadulterated survey frames (as outlined in \S\ref{sec:CFHT_data}). 

\begin{figure}
\begin{center}
\hbox{
\includegraphics[angle=0, viewport= 45 55 700 560, clip, width=\hsize]{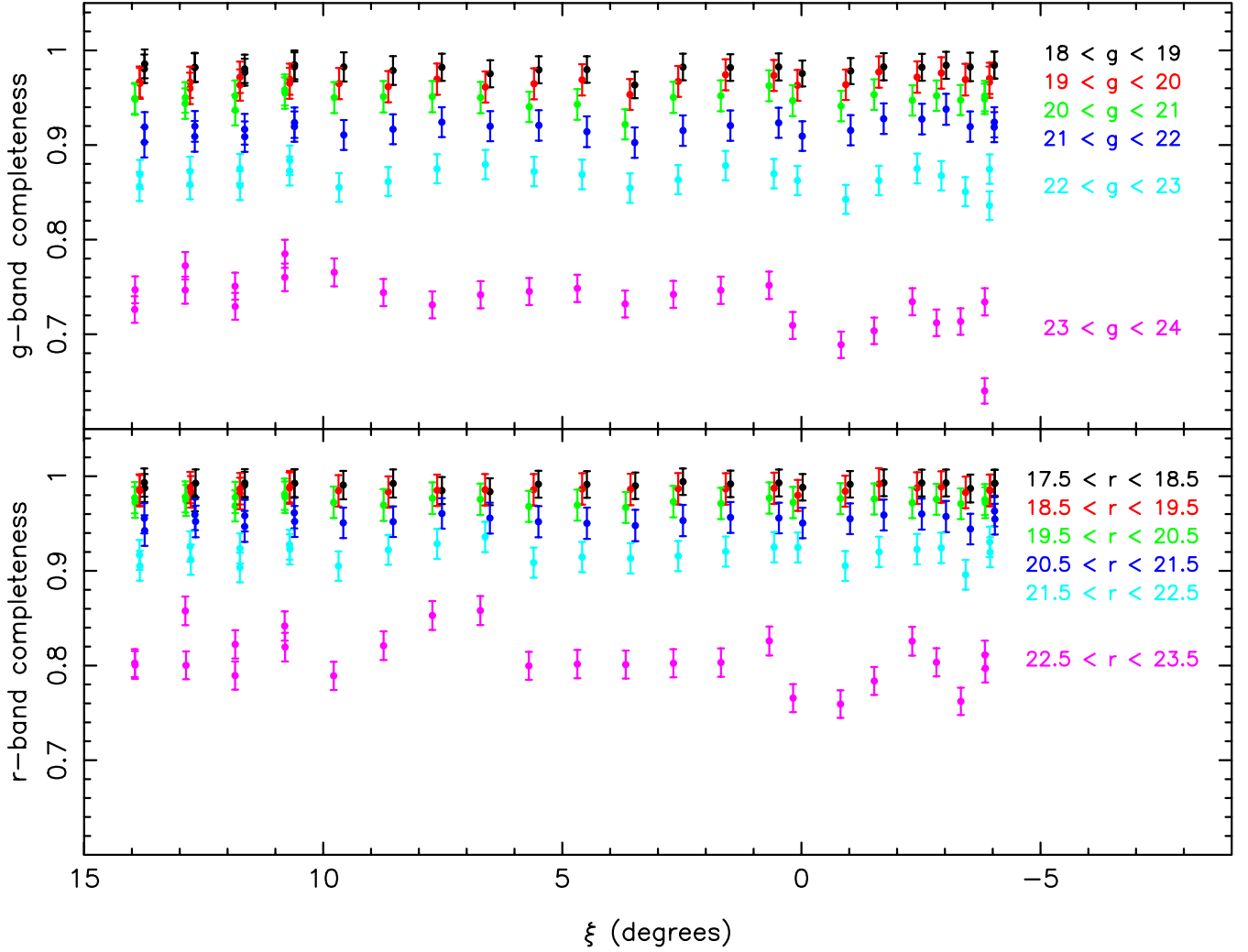}
}
\end{center}
\caption{Completeness fraction for the g-band (upper panel) and r-band data (lower panel). The magnitude intervals are shown in the same color as the corresponding measurements. For clarity, we have applied small offsets to the $\xi$ values in different magnitude intervals.}
\label{fig:completeness}
\end{figure}

Figure~\ref{fig:completeness} shows the resulting completeness as a function of position $\xi$ for stars of different magnitude. A star is considered to be recovered if a source is detected within $0\scnd5$ of the input position with a magnitude within 0.1~mags of the input value. The experiment shows that the stellar completeness is spatially highly uniform over the survey, which is not surprising since the survey covers sparse halo fields and the frames are of homogenous image quality.

\begin{table}
\caption{The first 10 rows of the CFHT/MegaCam catalog.}
\label{tab:CFHT}
\begin{tabular}{ccccccccccccccc}
\hline
\hline
$\alpha$ & $\delta$ & CCD & $x_{\rm g}$ & $y_{\rm g}$ & $\rm{g_0}$ & $\delta\rm{g}$ & $\rm{g_c}$ &
                            $x_{\rm r}$ & $y_{\rm r}$ & $\rm{r_0}$ & $\delta\rm{r}$ & $\rm{r_c}$ & field & E(B-V)\\
$(^{h~m~s})$ & $(\deg~'~'')$ & & & & & & & & & & & & & \\
\hline
 14:58:00.004 & -05:13:46.77 & 27 & 2124.31 & 2121.37 & 22.918 & 0.045 & -1 & 2123.29 & 2120.26 & 22.582 & 0.039 & -1 &  1 & 0.081\\
 14:58:00.098 & -04:59:34.04 & 18 &    8.00 & 2649.41 & 23.767 & 0.099 & -1 &    8.81 & 2650.18 & 22.702 & 0.045 & -1 &  1 & 0.092\\
 14:58:00.102 & -05:17:28.05 & 27 & 2119.05 &  929.78 & 23.895 & 0.106 & -1 & 2118.19 &  928.72 & 22.696 & 0.044 & -1 &  1 & 0.080\\
 14:58:00.117 & -05:03:36.17 & 18 &   15.23 & 3951.41 & 23.214 & 0.059 & -1 &   16.21 & 3952.08 & 23.020 & 0.058 & -1 &  1 & 0.089\\
 14:58:00.164 & -05:05:28.80 & 18 &   21.19 & 4556.88 & 21.739 & 0.017 & -1 &   22.25 & 4557.50 & 20.646 & 0.008 & -1 &  1 & 0.089\\
 14:58:00.184 & -04:09:39.00 & 27 & 2126.90 & 3438.39 & 24.158 & 0.102 & -2 & 2124.41 & 3436.78 & 22.697 & 0.045 & -2 &  2 & 0.104\\
 14:58:00.184 & -05:22:55.45 & 36 & 2118.70 & 4195.64 & 22.706 & 0.037 & -1 & 2117.54 & 4194.94 & 22.622 & 0.041 & -1 &  1 & 0.082\\
 14:58:00.254 & -04:01:20.08 & 18 &    7.39 & 3234.41 & 22.516 & 0.025 & -1 &    9.61 & 3235.70 & 22.168 & 0.029 & -1 &  2 & 0.117\\
 14:58:00.258 & -05:21:41.88 & 36 & 2111.11 & 4593.55 & 21.364 & 0.012 & -1 & 2110.03 & 4592.75 & 20.576 & 0.008 & -1 &  1 & 0.081\\
 14:58:00.262 & -05:01:01.24 & 18 &   23.56 & 3118.42 & 22.758 & 0.041 & -1 &   24.44 & 3119.16 & 21.636 & 0.018 & -1 &  1 & 0.092\\
\hline
\hline
\end{tabular}
\tablecomments{$\alpha$ and $\delta$ list the position of the star. The 3$^{\rm rd}$ column lists the CFHT CCD number on which the detection was made. Columns 4--8 provide the information on the g-band measures: $x_{\rm g}$ and $y_{g}$ are the position on the CCD, ${\rm g}$ and $\delta\rm{g}$ are the extinction-corrected magnitude and uncertainty, while $\rm{g_c}$ is a classification index. Columns 9--13 provide the same information for the r-band. Finally, column 14 lists the field number, and column 15 gives the dust reddening estimate. The classification index lists noise objects as 0, galaxies as 1, and stellar objects with negative integers, such that $-1$ corresponds to objects within $1\sigma$ of the stellar locus defined by the CASU software (see \citealt{Irwin:2001eq}), $-2$ corresponds to objects within $2\sigma$ of the stellar locus, etc. Objects with classification of $-9$ are saturated.}\end{table}

\begin{table}
\caption{The first 10 rows of the KPNO/Mosaic~II catalog.}
\label{tab:KPNO}
\begin{tabular}{ccccccccccccccc}
\hline
\hline
$\alpha$ & $\delta$ & ccd & $x_{\rm M}$ & $y_{\rm M}$ & $\rm{M_0}$ & $\delta\rm{M}$ & $\rm{M_c}$ &
                            $x_{\rm D}$ & $y_{\rm D}$ & $\rm{D_0}$ & $\delta\rm{D}$ & $\rm{D_c}$ & field & E(B-V)\\
$(^{h~m~s})$ & $(\deg~'~'')$ & & & & & & & & & & & & & \\
\hline
 15:00:06.746 & -04:31:46.54 &  1 & 1000.87 & 1097.23 & 13.632 & 0.001 & -9 & 1000.13 & 1094.48 & 13.317 & 0.001 & -9 &  1 &  0.106\\
 14:59:34.465 & -04:27:16.70 &  1 & 484.0   & 2024.35 & 13.857 & 0.001 & -9 & 482.03  & 2021.65 & 31.548 & 7.63  &  0 &  1 &  0.102\\
 15:00:38.820 & -04:31:17.19 &  1 & 935.05  & 164.89  & 14.078 & 0.001 & -9 & 886.74  & 759.15  & 13.842 & 0.001 & -9 &  1 &  0.109\\
 15:00:01.125 & -04:25:45.04 &  1 & 301.06  & 1256.94 & 14.237 & 0.001 & -9 & 301.03  & 1255.95 & 13.958 & 0.001 & -9 &  1 &  0.106\\
 14:59:34.453 & -04:27:13.53 &  1 & 477.84  & 2024.65 & 14.296 & 0.001 & -9 & 478.04  & 2024.02 & 13.01  & 0.001 & -9 &  1 &  0.102\\
 14:59:33.996 & -04:27:10.53 &  1 & 472.04  & 2037.85 & 14.448 & 0.001 & -9 & 472.03  & 2036.93 & 13.302 & 0.001 & -9 &  1 &  0.102\\
 14:59:43.184 & -04:23:20.29 &  1 & 21.09   & 1774.21 & 14.743 & 0.001 & -9 & 20.35   & 1771.79 & 14.836 & 0.001 & -9 &  1 &  0.104\\
 15:00:06.883 & -04:31:09.09 &  1 & 928.62  & 1093.26 & 15.166 & 0.001 & -9 & 927.92  & 1090.89 & 15.216 & 0.001 & -1 &  1 &  0.108\\
 15:00:05.117 & -04:24:59.17 &  1 & 210.17  & 1141.21 & 15.225 & 0.001 & -9 & 209.41  & 1138.98 & 15.298 & 0.001 & -1 &  1 &  0.107\\
 15:00:25.344 & -04:29:53.74 &  1 & 777.76  & 557.84  & 15.336 & 0.001 & -9 & 776.98  & 555.47  & 15.401 & 0.001 & -1 &  1 &  0.109\\
\hline
\hline
\end{tabular}
\tablecomments{As Table~\ref{tab:CFHT}, but for the KPNO catalog.}\end{table}

\end{document}